\documentstyle[preprint,12pt,aps]{revtex}

\begin{document}
\draft
\title{Low Temperature Phase of Asymmetric Spin Glass Model in Two Dimensions}
\author{T. Shirakura}
\address{Faculty of Humanities and Social Sciences, Iwate University, 
Morioka 020, Japan}

\author{F. Matsubara}
\address{Department of Applied Physics, Tohoku University, Sendai 980-77,
Japan}

\date{\today}
%
\maketitle
\begin{abstract}
We investigate low temperature properties of a random Ising model with 
$+J$ and $-aJ (a \neq 1)$ bonds in two dimensions using a cluster heat 
bath method. 
It is found that the Binder parameters $g_L$ for different sizes of the 
lattice come together at almost the same temperature implying the occurrence 
of the spin glass(SG) phase transition. 
From results of finite size scaling analyses, we suggest that 
the SG phase really occurs at low temperatures which is characterized 
by a power law decay of spin correlations.

\end{abstract}



Spin glasses have attracted  great challenge for computational physics in 
these two decades. It is widely believed now in the bond-random Ising model 
that spin glass (SG) transitions occur at a finite, non-zero 
temperature $T_c \neq 0$ in three dimensions(3D)\cite{BY0,OG0,OG1,SC} 
and at zero temperature $T_c = 0$ 
in two dimensions(2D)\cite{SC,MB1,BM1,BY1,Ozeki1}. 
Recently, the present authors\cite{SM1} reexamined the SG phase transition 
of the $\pm J$ Ising model on a square lattice of $L \times L$ by 
means of an exchange Monte Carlo method\cite{HN1} and found that the Binder 
parameters $g_L$ for $L \leq 16$ intersect at $T \neq 0$. 
They also found that better finite-size scaling(FSS) fits of the spin glass 
susceptibility $\chi_{SG}$ are obtained when $T_c \neq 0$. 
These results imply the occurrence of the SG phase transition at $T_c \neq 0$. 
If so, it is quite interesting, because it disproves the belief of $T_c = 0$.  
However, there remain two problems which should be considered 
to suggest $T_c \neq 0$ in 2D. 
One is that $g_L$ for a smaller lattice almost saturates below a rather high 
temperature\cite{BY1} and its saturation value slightly increases 
with $L$\cite{SM1}. 
Therefore, it is difficult to see whether the intersection of $g_L$ for 
smaller $L$ suggests the presence of the SG phase at $T_c \neq 0$ 
or is merely due to a finite size effect. 
The other is that it is still open whether or not the model really exhibits 
the nature of the SG phase at $T < T_c$, because the estimated transition 
temperature $T_c$ is a slightly lower than the lowest temperature 
which is reached in the simulation. 
The problems would be solved, if we study the same model on bigger lattices 
at lower temperatures.  
The saturation of $g_L$ at rather high temperatures, however, may be 
removed, if we treat an asymmetric random Ising model with $+J$ and 
$-aJ (a \neq 1)$ bonds, because the energy gap of $2|1-a|J$ in that model 
between the ground state and the lowest excitation state is much smaller than 
that of $4J$ in the $\pm J$ model\cite{COM1}, and, if the lattice is rather 
small, we may study equilibrium properties at any temperature 
using a cluster heat bath(CHB) method\cite{CHB1,CHB2}.

In this Letter, we investigate low temperature properties of 
the asymmetric random Ising model on the square lattice of 
$L \times L (L \leq 18)$ using the CHB method. 
In fact, $g_L$ does not saturate down to a very low temperature. 
We find that, as the temperature decreases, $g_L$'s for different $L$ 
meet at almost the same temperature and then increase together. 
This property rather resembles that of the $\pm J$ model in 3D 
in which the SG phase transition occurs at $T_c \neq 0$. 
We make the FSS analyses and find that $g_L$ and $\chi_{SG}$ for 
different $L$ scale well using a finite, non-zero value of $T_c$ 
and that the distribution functions $P_L(Q)$ of the spin overlap $Q$ scale 
at all temperatures below $T_c$.  
Thus we suggest that in this model the SG phase occurs at low temperatures 
which is characterized by a power law decay of the spin correlations. 
The properties for $T > T_c$ found here are very similar to those of 
the $\pm J$ model in 2D\cite{SM1}. 
We believe, hence, that the SG phase transition also occurs at $T_c \neq 0$ 
in the $\pm J$ model in 2D.


We start with an Ising model on a square lattice $L \times L$ described by 
the Hamiltonian
\begin{eqnarray}
     H = - \sum_{<i,j>}J_{ij}\sigma_{i}\sigma_{j}, 
\end{eqnarray}
where $\sigma_{i} (= \pm 1)$ are Ising spins and $<ij>$ runs all nearest 
neighbor pairs. Distributions of bonds for different samples are given as
\begin{eqnarray}
    P(J_{ij}) = \frac{1}{2}[\delta(J_{ij} - J) + \delta(J_{ij} + aJ)].
\end{eqnarray}
If the lattice is rather small and a free boundary condition
is used at least for one direction, we may readily obtain equilibrium spin 
configurations at any temperature $T$ by using the CHB 
method, because the cluster defined in Ref. 13 can be chosen as the 
lattice itself. That is, the cluster is composed of $L$ layers with $L$ spins
and exchange fields from the outside are absent. 
We briefly note the method\cite{COM2}. 
For every sample, the weight functions $F_l(\{\sigma_i^{(l)}\})$ can be 
uniquely determined by using Eq.(7) in Ref. 13, because $h_i^{(l)} = 0$. Once 
the set of these functions $\{ F_l(\{\sigma_i^{(l)}\}) \}$ is obtained, the
spin configurations of individual layers can be determined successively from 
the layer $(L)$ to the layer $(1)$ by using Eq.(8) or Eq.(11) in Ref. 13 
and random numbers. 
Thus one of the spin configurations of the lattice is generated. 
Repeating this procedure, we may generate any number of the spin 
configurations  with the aid of $\{ F_l(\{\sigma_i^{(l)}\}) \}$. 
These spin configurations are independent with each other and 
in accordance with the Boltzmann's weight\cite{CHB2}. 
For each of the samples, about $M = 200$ spin configurations are 
generated\cite{COM3}. 
We calculate, as well as usual magnetic quantities, an overlap function of 
the spins between the different spin configurations, $P_L^{(k)}(Q)$, 
for every sample: 
\begin{eqnarray}
  P_L^{(k)}(Q) = \frac{2}{M(M-1)}\sum_{n}^{M}\sum_{m(>n)}^{M}
\delta(Q - Q_{nm}^{(k)}) 
\end{eqnarray}
with $Q_{nm}^{(k)} = (1/N)\sum_{i=1}^N\sigma_i^{(n,k)}\sigma_i^{(m,k)}$, where 
$\sigma_i^{(n,k)}$ is the $i$th spin of the $n$th spin configuration for the 
$k$th sample. The overlap function $P_L(Q)$ of the system is given as 
$P_L(Q) = (1/N_s)\sum_{k=1}^{N_s}P_L^{(k)}(Q)$, where 
$N_s$ is the number of the samples. 
Once $P_L(Q)$ is determined, we may obtain various SG quantities. 
The $n$th moment of the spin overlap is defined as:
\begin{eqnarray}
  [<Q^n>] = \int_{-1}^{+1}Q^n P_L(Q) dQ, 
\end{eqnarray}
where $<\cdots>$ and $[\cdots]$ mean the spin configuration(thermal) average 
and the sample average, respectively. 
The SG susceptibility $\chi_{SG}$ is determined from
\begin{eqnarray}
  \chi_{SG} = N[<Q^2>], 
\end{eqnarray}
and the Binder parameter $g_L$ from
\begin{eqnarray}
  g_L = (3-[<Q^4>]/[<Q^2>]^2)/2. 
\end{eqnarray}
We have performed this CHB simulation of the model of Eq.(1) with 
$a = 0.8$ for $L \leq 18$. 
The numbers of the samples are $N_s = 4000$ for $L \leq 14$ and 
$N_s = 1000$ for $L = 16$ and $18$.

Figure 1 shows $P_L(Q)$ for different $L$. For every size $L$, 
the shape is symmetric with respect to $Q = 0$ and the peaks at 
$Q \sim \pm 1$ become steeper as the temperature decreases. 
Figure 2 shows plots of $g_L$ against $T$. As the temperature decreases, 
$g_L$'s for different $L$ meet at almost the same temperature of 
$T \sim 0.2J$ and then increase together. 
This behavior is quite similar to that of $g_L$ of the $\pm J$ model 
in 3D in which the SG phase transition occurs at $T_c \neq 0$\cite{BY0}.

We examine the results in more detail using the FSS analyses. 
First, we perform the FSS plots of $g_L$ and $\chi_{SG}$ to estimate the 
value of $T_c$.  
If the SG transition occurs at $T_c$, $g_L$ and $\chi_{SG}$ scale as 
\begin{eqnarray}
	g_L &=& G(\epsilon L^{1/\nu }),  \\
	\chi _{SG} &=& L^{2-\eta }X(\epsilon L^{1/\nu }),  
\end{eqnarray} 
where $\epsilon =(T-T_c)/J$, $\nu$ is the exponent of the correlation length, 
$\eta$ is the exponent which describes the decay of the spin 
correlation at $T=T_c$, and $G$ and $X$ are some scaling functions. 
Having assumed $T_c \simeq 0.19J$, we could obtain good scaling plots 
for $T \geq T_c$. Typical examples are shown in Figs. 3(a) and 3(b). 
Of course, the values of $\nu$ estimated from both the scaling plots 
are almost the same. 
We also examined the other possibility of $T_c = 0$. As for 
$\chi_{SG}$, we could scale the data only in the neighborhood of $T = 0$ 
using $\eta \sim 0$ and $\nu \sim 2.6$. 
As for $g_L$, on the other hand, we could never scale the data using 
any plausible value of $\nu$. 
The scaling plots for $T_c = 0$ are shown in Figs. 4(a) and 4(b). 
These results clearly reveal that, if a conventional phase transition 
occurs, the transition temperature is $T_c \sim 0.19J$, not $T_c \sim 0$. 
However, the data for $T < T_c$ deviate from the scaling plots.

Next we examine $P_L(Q)$ itself to see whether or not the SG phase 
is realized below $T < T_c$. 
If the phase transition occurs at $T = T_c$, $P_L(Q)$'s for different $L$ 
will scale as 
\begin{eqnarray}
    P_L(q) &=& L^{\eta /2}P(qL^{\eta /2}) \;\;\;\; {\rm at} \;\;\;\;  T=T_c . 
\end{eqnarray}
Since $P_L(Q)$ for smaller $L$ has a considerable weight at $Q = \pm 1$ 
and, for $L \leq 16$, the peak height rather decreases with increasing $L$, 
we could not scale all the data over the entire range of $Q$. 
These difficulties may, however, come from finite size effects.  
In fact, the weights at $Q = \pm 1$ become smaller as $L$ increases 
and, for $T \leq 0.2J$, the peak height for $L = 18$ becomes larger 
than that for $L = 16$.  
If we overlook the discrepancy around the peak of $P_L(Q)$, the data 
scale for $L \geq 10$ at $T \sim T_c$  by using the same value of 
$\eta \sim 0.14$ for $\chi_{SG}$, which is shown in Fig. 5(a).  
However, the data for $T < T_c$ can also be scaled by using 
a smaller value of $\eta$ as shown in Fig. 5(b)\cite{COM4}. 
A natural interpretation of this result is that the model is close 
to criticality at all temperatures below $T_c$ like 
the $XY$ ferromagnet in 2D\cite{KT}. 
This picture is, of course, compatible with the fact that $g_L$ 
and $\chi_{SG}$ scale only for $T \geq T_c$. 
We suggest, hence, that the SG phase really occurs below $T_c \sim 0.19J$ 
which 
is characterized by a power law decay of the spin correlations.

Our present result is in agreement with our previous result of $T_c \neq 0$ 
in the $\pm J$ model\cite{SM1}. 
Especially, the value of $\nu \sim 0.18$ in the $\pm J$ model is in good 
agreement with that obtained in the present model\cite{COM5}. 
Thus we predict that the occurrence of the SG phase at $T_c \neq 0$ is the 
common nature of 2D random Ising models with a discrete distribution of 
bonds\cite{COM6}. 
Our prediction of $T_c \neq 0$ appears incompatible with the previous belief 
of $T_c = 0$. 
We think, however, that these are not necessarily incompatible, 
because the previous authors have only concluded that {\it their data are not 
incompatible with the prediction that $T_c = 0$, and have not ruled out 
the possibility of such low $T_c$ as estimated here}. 
The thing that was certainly suggested by the previous studies is that, 
at $T = 0$, no long-range order exists and
the spin correlation decays according to the power law\cite{MB1,BM1,Ozeki1}. 
This is compatible with our result with $T_c \neq 0$. 
Of course, further studies are necessary to confirm the prediction 
of $T_c \neq 0$. 
We believe that the present result will stimulate not only the computational 
physics but also experimental studies, because the bond distributions of
real SG materials will be asymmetric.

\bigskip

One of the authors (TS) wishes to thank Professor H. Takayama, Professor
H. Kawamura, Dr. Y. Ozeki, Dr. K. Hukushima and Dr. H. Yoshino 
for valuable discussions. 
The simulations were made partly on FACOM VPP500 at the Institute for 
Solid State Physics in University of Tokyo.



\begin{figure}
\caption{$P_L(Q)$ versus $Q$ at (a) $T=0.3J$, (b) $T=0.2J$ and (c)
$T=0.1J$ in the random Ising model with $a=0.8$ on the square lattice of
$L \times L$.}
\end{figure}

\begin{figure}
\caption{Temperature dependences of $g_L$ of the random Ising model with
$a=0.8$.}
\end{figure}

\begin{figure}
\caption{Scaling plots of (a) $g_L$ and (b) $\chi_{SG}$, assuming
$T_c=0.19J$ and $\epsilon =(T-T_c)/J$.}
\end{figure}

\begin{figure}
\caption{Scaling plots of (a) $g_L$ and (b) $\chi_{SG}$, assuming
$T_c=0$ and $\epsilon =(T-T_c)/J$.}
\end{figure}

\begin{figure}
\caption{Scaling plots of $P_L(Q)$ at (a) $T=0.2J$ and (b) $T=0.1J$.}
\end{figure}


\begin{references}
\bibitem{COM1}  But the number of the ground states reduces drastically 
in the asymmetric model, which will be reported elsewhere.
\bibitem{COM3} We have examined the accuracy of results comparing data 
averaged over $M$ spin configurations with those over 
$M/2$ spin configurations and found that the difference is less than 0.1$\%$ 
after sample averages are taken out.
\bibitem{COM4} These values of $\eta$ are consistent with those estimated 
from the slope of the log-log plots of $\chi_{SG}$ versus $L$.
\bibitem{COM6} At present, we could not give the same prediction for the 
model with a Gaussian distribution of bonds, because the distribution 
of excitation energies of that model is quite different from that of the 
present model.
\end{references}
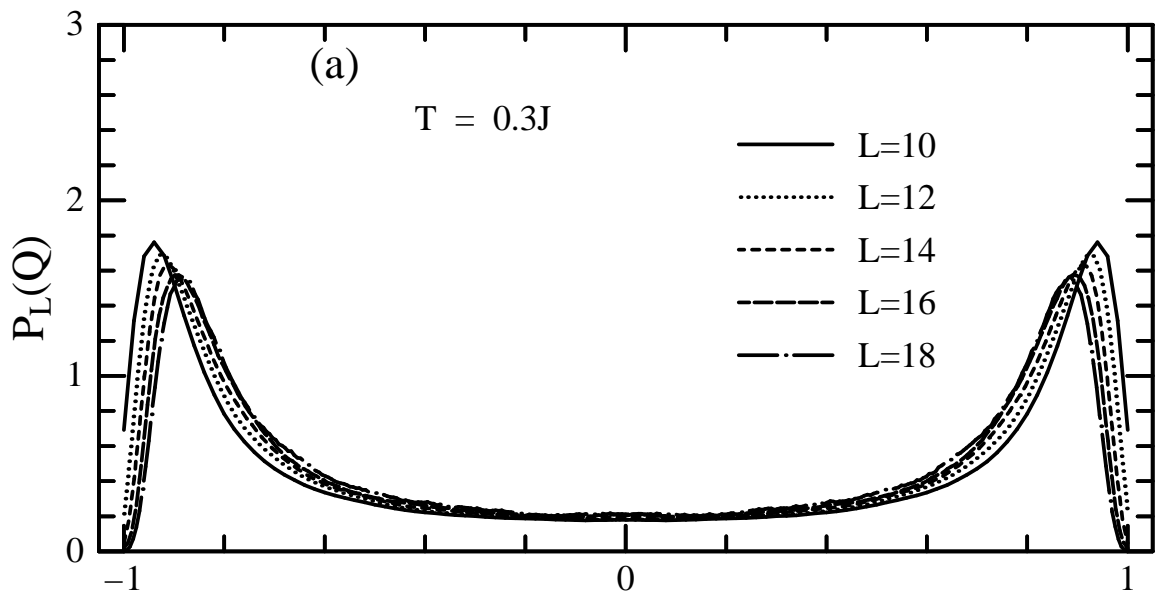
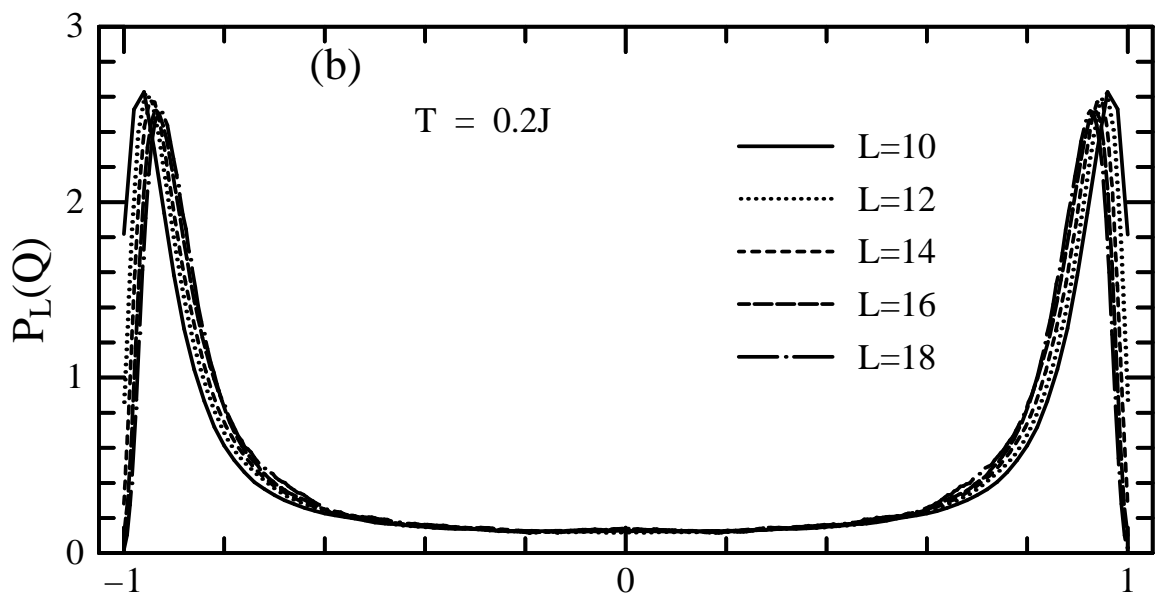
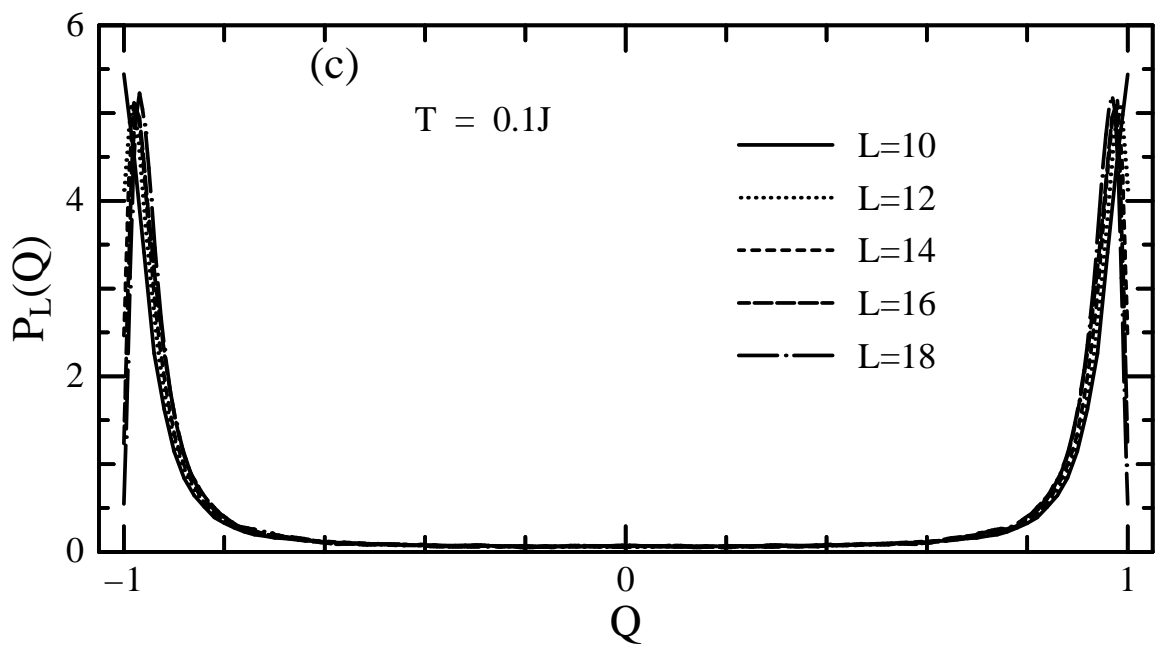
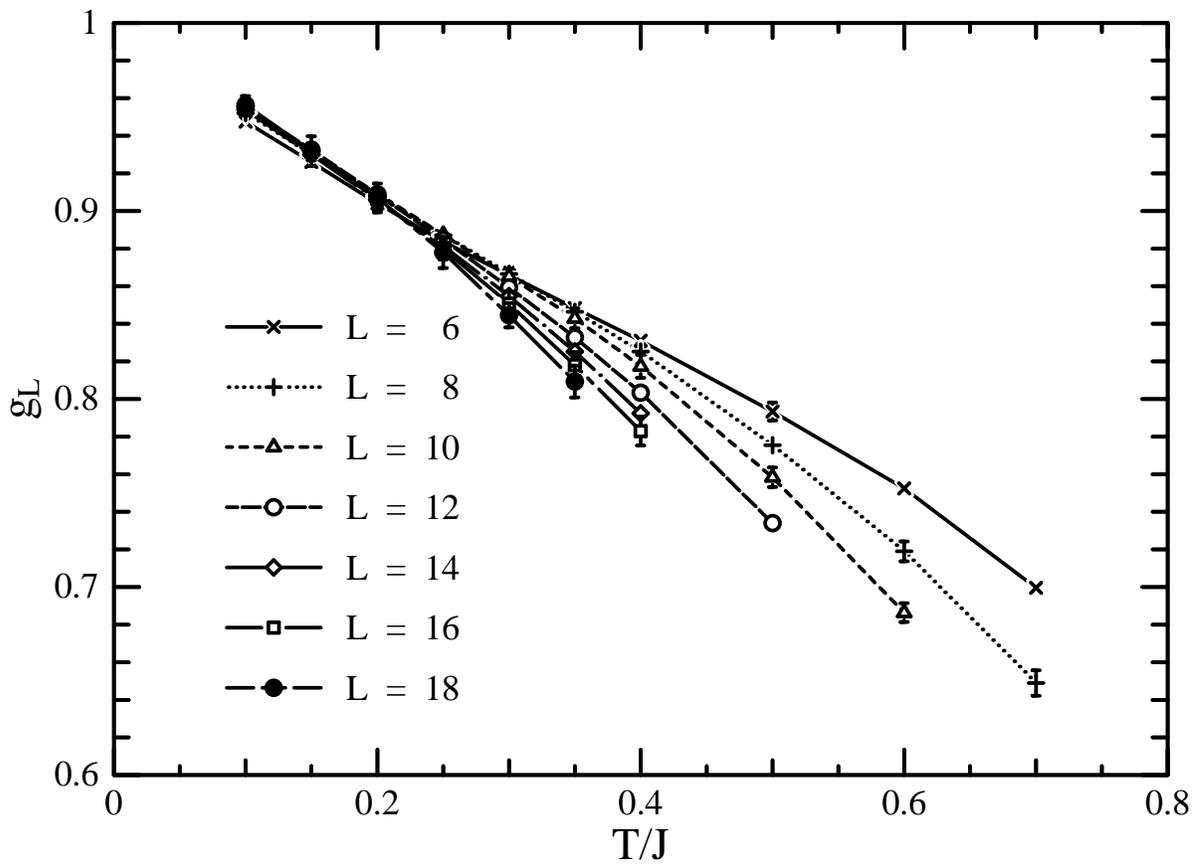
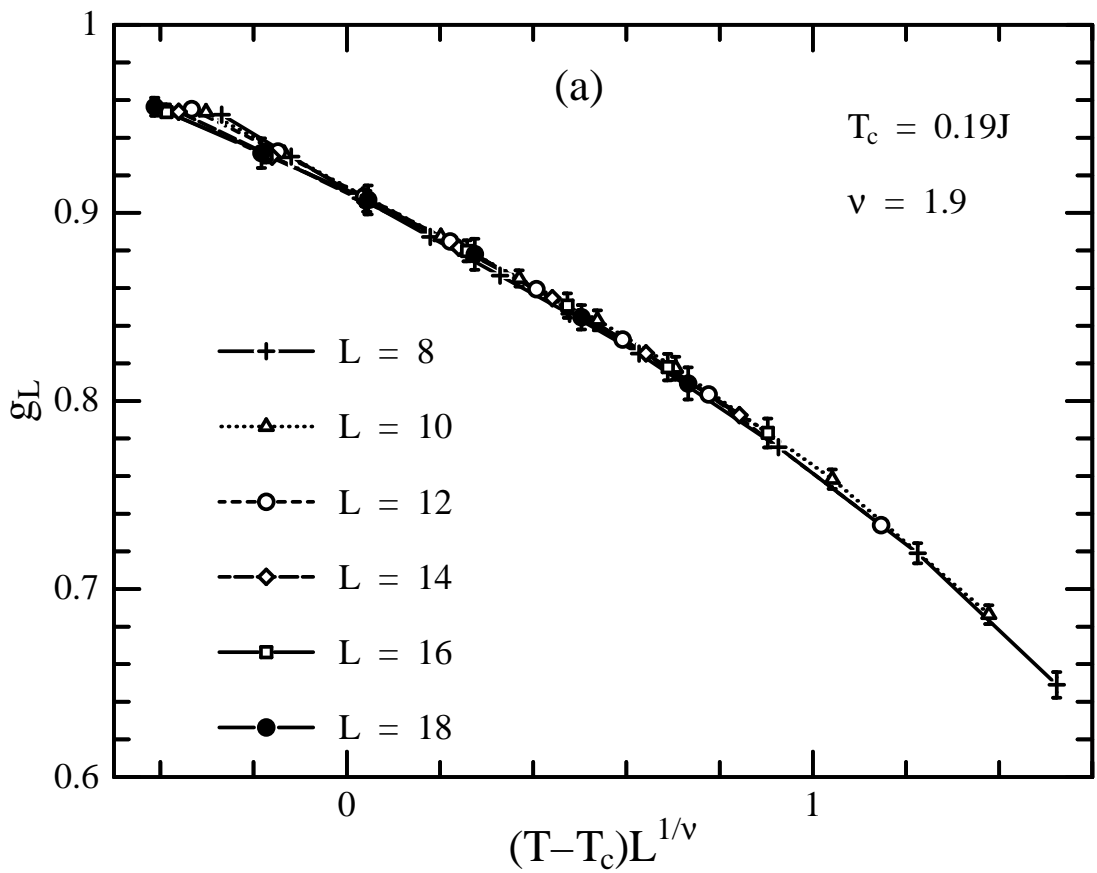
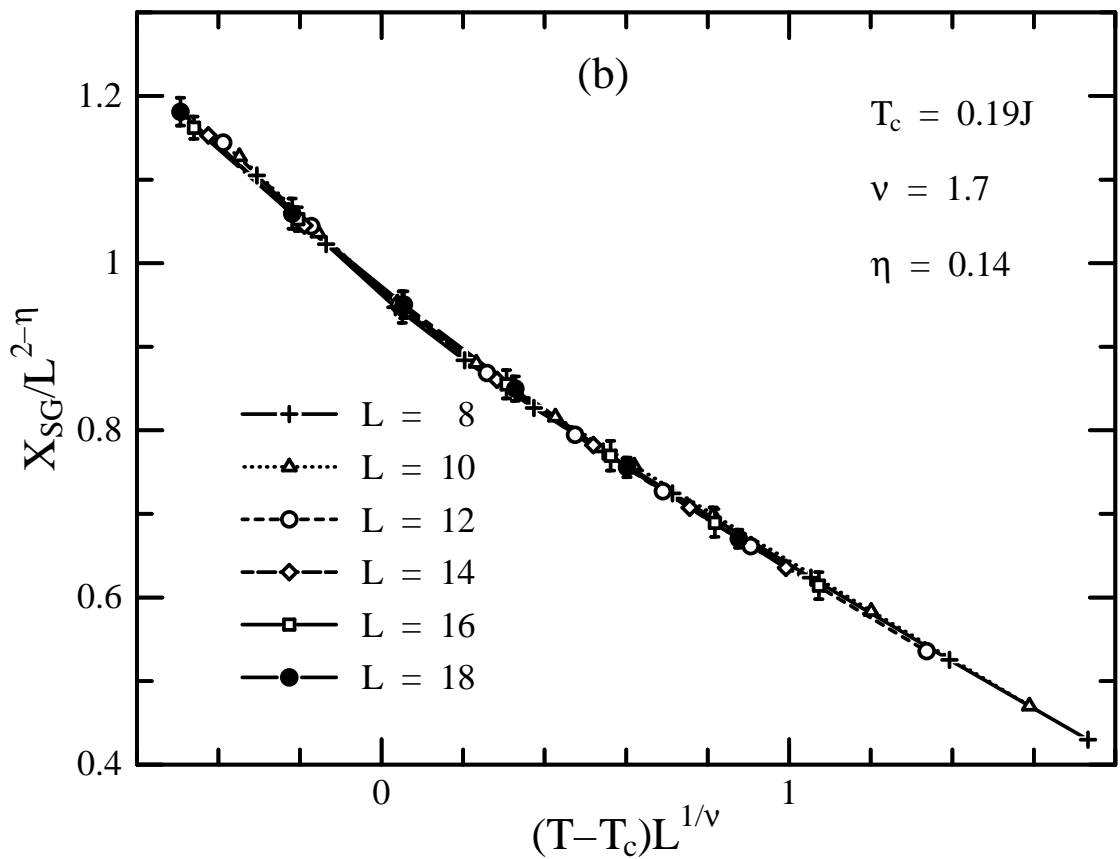
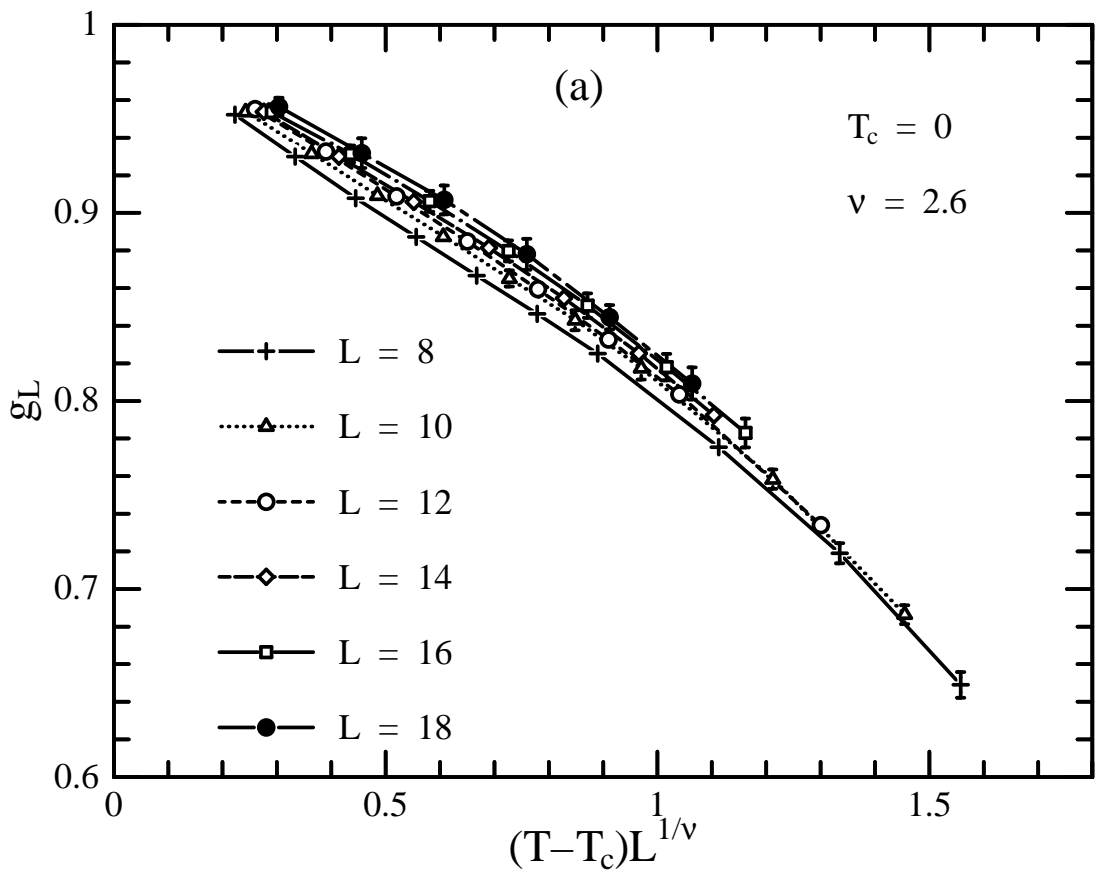
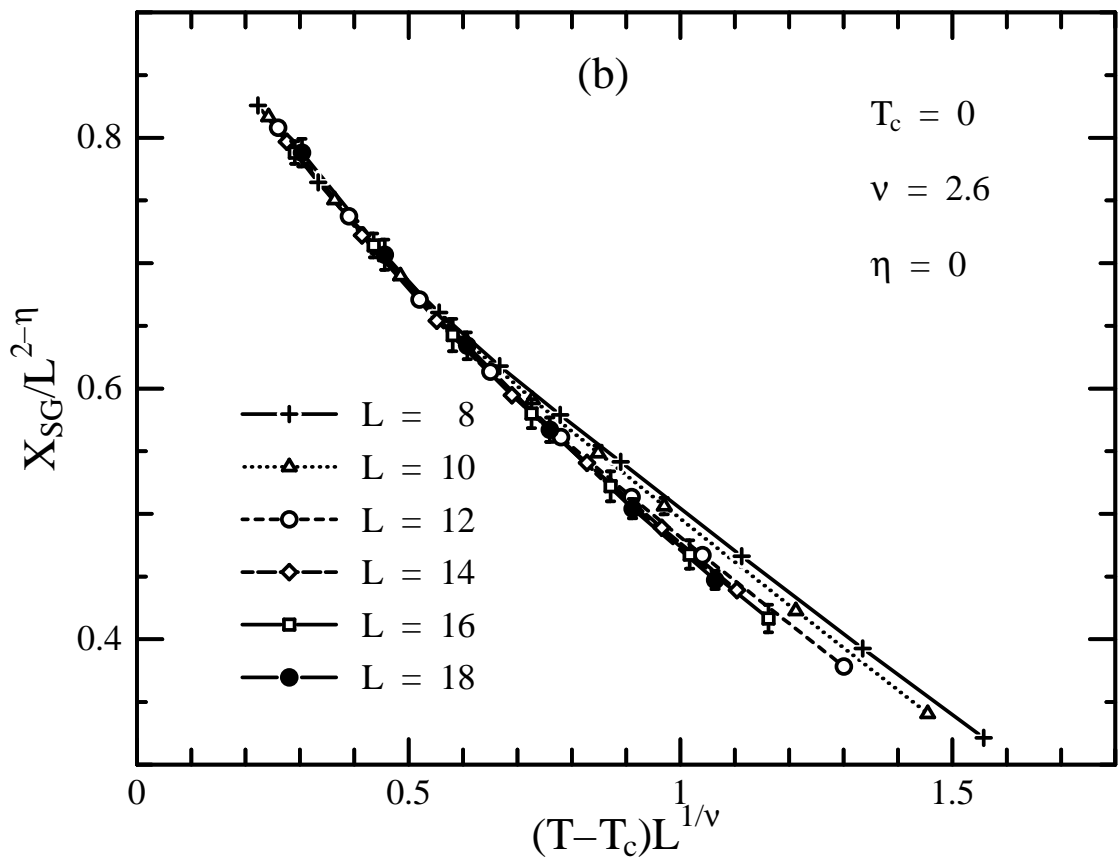
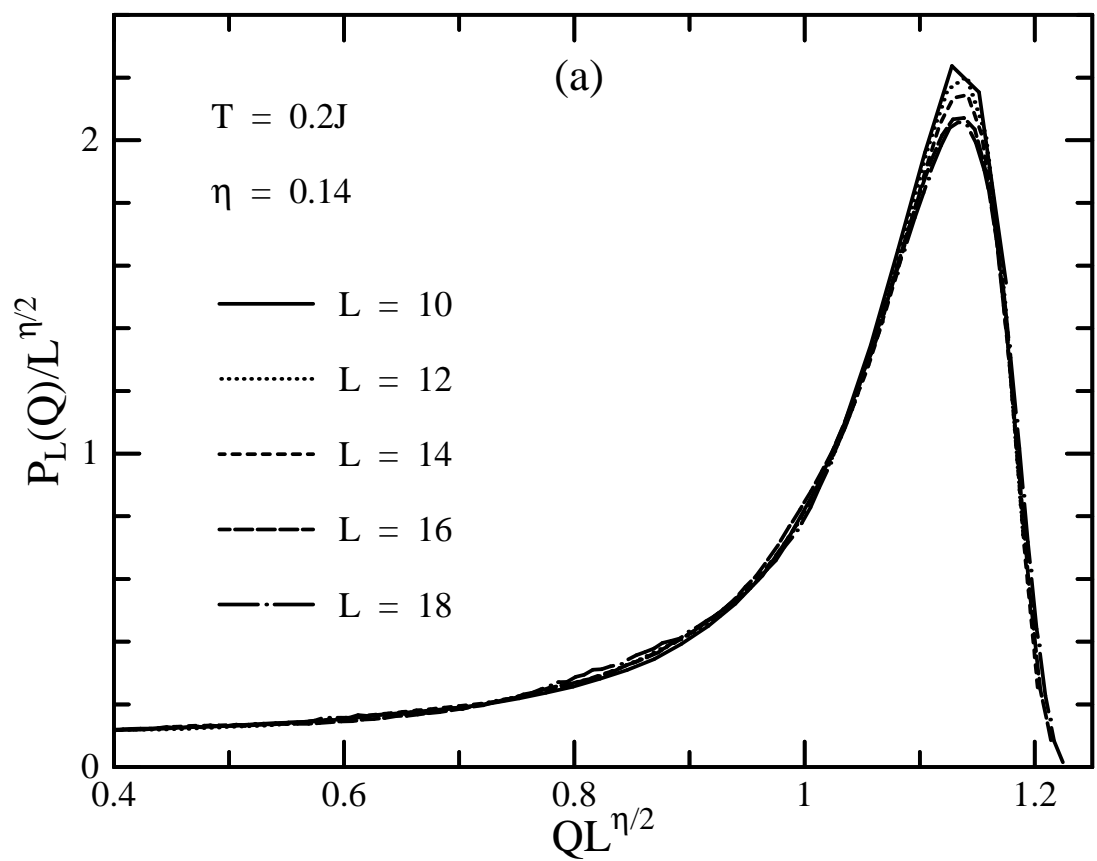
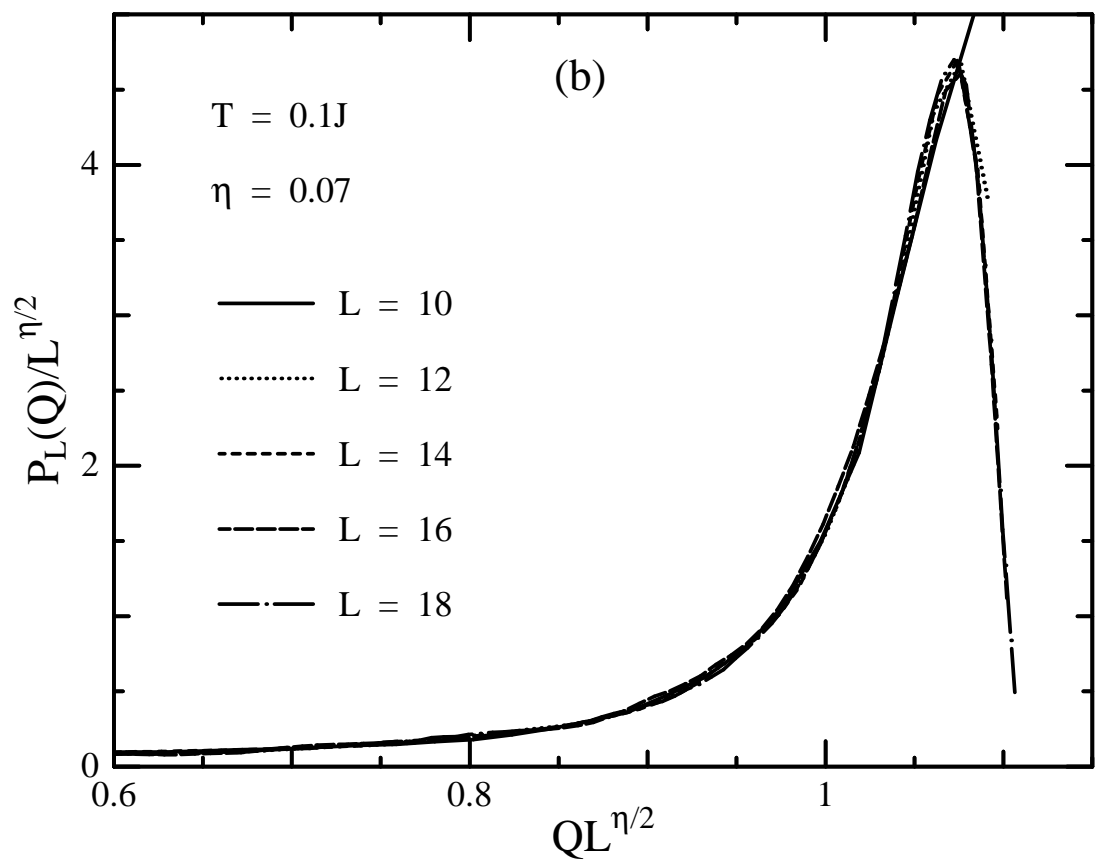
\end{document}